\documentstyle[12pt,epsfig]{article}

\begin{document}
\topmargin -1.4cm
\oddsidemargin -0.8cm
\evensidemargin -0.8cm 

\title{Inverse transfer of self-similar decaying turbulent\\ 
non-helical magnetic field}

\vspace{1.5cm}

\author{P. Olesen\\
{\it  The Niels Bohr Institute}\\
{\it Blegdamsvej 17, Copenhagen \O, Denmark }}

\maketitle

\begin{abstract}
We show that decaying turbulent non-helical magnetic fields satisfy a 
self-similarity relation according to which the relevant scales increase as 
time passes (inverse cascade or inverse transfer). We compute 
analytically quantities which have previously been determined by numerical 
calculations, for example the average energy and the
integral scale which are proportional to $1/t$ and $\sqrt{t}$, respectively,
where $t$ is time. We also briefly discuss self-similarity for the helical
case.
\end{abstract}

\vskip0.5cm

The important subject of freely decaying non-helical magnetic fields have 
recently caught much interest \cite{axel},\cite{zrake}. These fields are of 
interest for cosmology (for a recent review, see \cite{axel2}) and 
for the interpretation of new
measurements of optical polarization in relation to gamma ray burst
afterglows (for a recent discussion, see \cite{gamma}). The interesting 
feature of the investigations \cite{axel} and \cite{zrake} is that there
is an inverse transfer (also called an inverse cascade\footnote{Some 
people consider this terminology
to be politically incorrect and it should be reserved for
the helical case only.}) from small to large 
scales of the magnetic field, in spite of the
fact that there is no magnetic helicity. 

In this note we shall show that some of the results obtained numerically in
refs. \cite{axel} and \cite{zrake} can be understood analytically quite 
rigorously in  magnetohydrodynamics (MHD). We shall find that the
kinetic and magnetic energy spectra are self-similar, as found numerically 
for the magnetic field energy by 
Zrake\footnote{His calculation is done in relativistic MHD, whereas our result
refers to non-relativistic MHD} \cite{zrake}. We start by giving the main
result,
\begin{equation}
{\cal E}_{v,B}(k,t)=\sqrt{\frac{t_0}{t}}~{\cal E}_{v,B}
\left(k\sqrt{\frac{t}{t_0}},t_0\right),
\label{stuff}
\end{equation} 
where ${\cal E}_{v,B}$ are the kinetic and magnetic energy densities 
normalized as
\begin{equation} 
\int_0^\infty dk ~{\cal E}_v=\frac{1}{2}~<{\bf v}({\bf x},t)^2>,~  
\int_0^\infty dk~ {\cal E}_B=\frac{1}{2}~\left<{\bf B}({\bf x},t)^2\right>
\label{2}
\end{equation}
for isotropic turbulence. The scaling exhibited in Eq.(\ref{stuff}) is
expanding in the sense that the effective $k$ is shifted towards $k=0$,
i.e. towards larger distances. 

Eq. (\ref{stuff}) is valid when the unforced MHD equations are satisfied. 
If the
initial conditions are established by some initial forcing (``stirring'') 
then the 
self-similarity (\ref{stuff}) only sets in after the switching off of the
external forces. The actual form of the initial conditions is of no relevance
for the self-similarity. Our result is valid for the non-relativistic version
 of the MHD equations. This may explain a slight discrepancy between Eq.
(\ref{stuff}) and
the result in ref. \cite{zrake}. The prefactor $\sqrt{t_0/t}$ in 
(\ref{stuff}) is slightly different from the one obtained by Zrake, and the
power of $t/t_0$ associated with $k$ is also slightly different.

From (\ref{stuff}) we immediately obtain the
time dependence of the integrated energy density (we assume that $\cal E$ is 
not a constant)
\begin{equation}
{\cal E}_{v,B}(t)=\int_0^\infty dk~{\cal E}_{v,B}(k,t)=
\frac{t_0}{t}\int_0^\infty dx~{\cal E}_{v,B}(x,t_0)\propto \frac{1}{t}.
\end{equation}
This is in accordance with the numerical results obtained by Kahniashvili, 
Brandenburg, and  Tevzadze \cite{axel}. As we shall see from the derivation of
(\ref{stuff}) this result is rigorous and is thus a also a check of 
the numerical
calculations. Similarly we can obtain the kinetic or magnetic integral scale 
\begin{equation}
<\xi>=\kappa^{-1}(t)=\frac{\int_0^\infty~\frac{dk}{ k}~{\cal E}(k,t)}
{ {\cal E}(t) }=\sqrt\frac{t_0}{t}~\frac{\int_0^\infty\frac{dx}{x}
{\cal E}(x,t_0)}{{\cal E}(t)}\propto \sqrt{t},
\end{equation}
again in accordance with ref. \cite{axel}. Higher moments can be computed,
\begin{equation}
<\xi^2>\propto t,~~~    <\xi^3>\propto t^{3/2},
\end{equation}
etc. For the higher moments there may be convergence problems at $x=0$, so it
may be more productive to consider moments 
\begin{equation}
<k^n>\propto t^{-n/2}.
\end{equation}

The derivation of the main equation (\ref{stuff}) follows to some extent 
an earlier paper
\cite{poul} (see also \cite{peter}), except that we do not attempt to 
interpret the results in terms
 of initial values. The latter required some assumptions on inertial 
range behaviour.
A critical discussion of problems in this connection is given in refs.
\cite{axel} and \cite{axel3}.

To get (\ref{stuff}) we start from the observation that the unforced
MHD equations 
\begin{equation}
\partial_t{\bf v}+{\bf (v\nabla )v}=-\nabla p+
{\bf (\nabla\times B)\times B}+\nu
\nabla^2{\bf v},~~~\partial_t{\bf B}={\bf \nabla\times(v\times B)}+\eta
{\bf \nabla^2B},
\end{equation}
are invariant under the scalings
\begin{equation}
{\bf x}\rightarrow l{\bf x},~t\rightarrow l^2t,~{\bf v}\rightarrow {\bf v}/l,
{\bf B}\rightarrow {\bf B}/l,~\nu\rightarrow \nu,~\eta\rightarrow \eta,~p
\rightarrow p/l^2.
\label{scale}
\end{equation}
The crucial point is that the kinetic and Ohmic diffusions $\nu$
end $\eta$, respectively, are invariant under this scaling. Let us consider 
the energy density
\begin{equation}
{\cal E}_v(k,t)=\frac{2\pi k^2}{(2\pi)^3}~\int d^3y~e^{i{\bf k y}}
< {\bf v}({\bf x},t){\bf v}({\bf x+y},t)>,
\label{begin}
 \end{equation}
with a similar expression for the magnetic energy density with $\bf v$ replaced
by $\bf B$ on the right hand side. It is easily seen that integration of both
sides of Eq. (\ref{begin}) gives Eq. (\ref{2}). We now replace the variables
according to the scalings (\ref{scale}) to obtain (${\bf y}=l{\bf y'}$)
\begin{equation}
{\cal  E}_v (k/l,l^2t)=l~\frac{2\pi k^2}{(2\pi)^3}~\int d^3y'~
e^{i{\bf k y'}}< {\bf v}(l{\bf x'},l^2t){\bf v}(l{\bf (x'+y')},l^2t)>
=l^{-1}{\cal E}_v(k,t),
\end{equation}
with a similar expression for ${\cal E}_B$. Thus
\begin{equation} 
{\cal  E}_{v,B} (k,t)=l{\cal  E}_{v,B} (k/l,l^2t)
\label{x}
\end{equation}
This is a functional relation for all values of the parameter $l$. 
We want now to compare the situation for 
two different times $t_0$ and $t$ by taking
\begin{equation}
l=\sqrt{t_0/t},
\end{equation}
leading immediately to Eq. (\ref{stuff}). Here all considerations are valid 
for the full interval of $k$.

In an earlier paper \cite{poul} we solved Eq. (\ref{x}) by
\begin{equation}
{\cal  E}_{v,B} (k,t)=k\psi (k^2t),
\end{equation}
which can be brought in the form of Eq. (\ref{stuff}) by writing
\begin{equation}
k^2t=(k\sqrt{t/t_0})^2t_0.
\end{equation}
The former approach \cite{poul},\cite{peter} also involves the use of 
the invariance of 
the MHD equations under the scalings
\begin{equation}
{\bf x}\rightarrow l{\bf x},~t\rightarrow l^{1-h}t,~{\bf v}\rightarrow 
l^h{\bf v},
{\bf B}\rightarrow l^h{\bf B},~\nu\rightarrow l^{1+h}\nu,~\eta\rightarrow 
l^{1+h}\eta.
\label{super}
\end{equation}
This scaling is physically somewhat unpleasant to apply for $h\neq -1$
since it involves changing diffusions,
i.e. moving between different universes. For further critical discussions
of the approach in \cite{poul} and \cite{peter} 
we refer to refs.\cite{axel} and \cite{axel3}. In particular the
identification of the time $t=0$ as the initial time is doubtful because
this scaling is not valid when the process of turbulence is initiated
by some forces. Also $t=0$ may be a singular time \cite{axel3}.

Similar critical remarks and reinterpretations of ref.\cite{poul} were made 
by Campanelli
\cite{campanelli}. He derived the result that for constant dissipation
 parameters the energy decays as $1/t$ and the integral scale goes like
$\sqrt{t}$. He also found  cosmological uses of the case where the
diffusions change as in the scaling Eq. (\ref{super}).

The physical picture which emerges from the self-similarity (\ref{stuff})
is that as time passes the physical scales increase. Any structure of size
$L$ at time $t_0$ will expand into $\sqrt{t/t_0} L$ at time $t$. 
This is similar to what happens in 
the expanding universe, with the factor $\sqrt{t/t_0}$ being analogous to the
scale factor in cosmology. The self-similar expansion is basically
inherent in the 
dynamical properties of the  MHD equations even in flat space. As discussed 
alreay in refs. \cite{poul} and \cite{campanelli} this can be extended
quite simply to the expanding universe. In general the inverse transfer makes 
the decay of the magnetic field slower than dictated by the expansion of the 
universe itself. This may be a more general phenomenon applicable to other 
vector fields, thereby making the conventional (text book) wisdom on the
rapid decrease of vector fields relative to scalars somewhat insecure. Of 
course, a detailed 
analysis is needed for other vector fields in each case in order to see if 
they have inverse
transfer.

Although we have considered the non-helical case it may be possible to
generalize to the case where we have helicity, provided this kind of 
turbulence is really statistically isotropic, which may be somewhat doubtful.
Assuming self-similarity we have in analogy with the
construction in Eq.(\ref{begin}) 
\begin{equation}
{\cal H}(k,t)=\frac{4\pi k^2}{(2\pi)^3}~\int d^3y~e^{i{\bf k y}}
< {\bf A}({\bf x},t){\bf B}({\bf x+y},t)>.
\end{equation}
Scaling now gives
\begin{equation}
{\cal H}(k,t)={\cal H}(k\sqrt{t/t_0},t_0)
\label{hhh}
\end{equation}
In ideal MHD with $\eta=0$ the helicity is constant. However, in the
presence of diffusion helicity decays. With
\begin{equation}
H(t)=\int_0^\infty dk~~{\cal H}(k,t)=<{\bf A}({\bf x},t){\bf B}({\bf x},t)>
\end{equation}
we have
\begin{equation}
H(t)=\int_0^\infty dk {\cal H}(k,t)\propto 1/\sqrt{t}.
\end{equation}
Thus helicity decreases slower with time than the energy $\propto 1/t$.
As observed in ref. \cite{axel} there will always be some helicity, due to 
fluctuations. These should then follow the self-similarity (\ref{hhh}).

As mentioned before self-similarity may not be right for the helical
case, especially if one considers long time intervals where $H$ is constant.
As an example there are the results in ref. \cite{bis} according to which
the energy decays as $1/\sqrt{t}$ for the helicity $H$ fixed. Such a behavior 
is clearly not covered by our results. A  discussion of the
problem of helicity versus scaling has been given by 
Campanelli \cite{campanelli}.

The conclusion from this note is that the inverse transfer reported in the
literature can in fact be understood as a result of the well known scaling 
relations (\ref{scale}), and the decay of the energy and the behavior
of the integral scale
as $1/t$ and $\sqrt{t}$, respectively, are simple consequences of the 
self-similarity (\ref{stuff}) resulting from these scaling relations. The
inverse transfer is therefore a generic feature of freely decaying
(magneto-)hydrodynamics.

\vskip.3cm

I thank Leonardo Campanelli for some comments.

%\vskip.3cm

\end{document}